\documentclass[letterpaper, 10 pt, conference]{ieeeconf}

\IEEEoverridecommandlockouts
\usepackage{epsfig}

\title{\LARGE \bf

Counterfactual Explanation-Based Badminton Motion Guidance Generation Using Wearable Sensors
}

\usepackage{amsmath} 
\usepackage{graphicx} 
\usepackage{multirow}
\usepackage{lineno}
\usepackage{tikz}
\usepackage{booktabs}
\usepackage{algorithm}
\usepackage{algorithmicx, algpseudocode}
\usepackage{soul}
\usepackage{comment}
\usepackage{cuted}

\author{Minwoo Seong$^{1}$, Gwangbin Kim$^{1}$, Yumin Kang$^{1}$, Junhyuk Jang$^{1}$, Joseph DelPreto$^{2}$, and SeungJun Kim$^{1*}$
\thanks{$^{1}$Minwoo Seong, Gwangbin Kim, Yumin Kang, Junhyuk Jang, and SeungJun Kim are with the School of Integrated Technology, Gwangjun Institute of Science and Technology, Gwangju, South Korea; *SeungJun Kim is the corresponding author.
$^{2}$Joseph DelPreto is with the Computer Science and Artificial Intelligence Laboratory, Massachusetts Institute of Technology, Cambridge, Massachusetts, USA;}
}

\begin{document}

\maketitle
\thispagestyle{empty}
\pagestyle{empty}

\begin{figure*}[t]
\centering
\includegraphics[width=\linewidth]{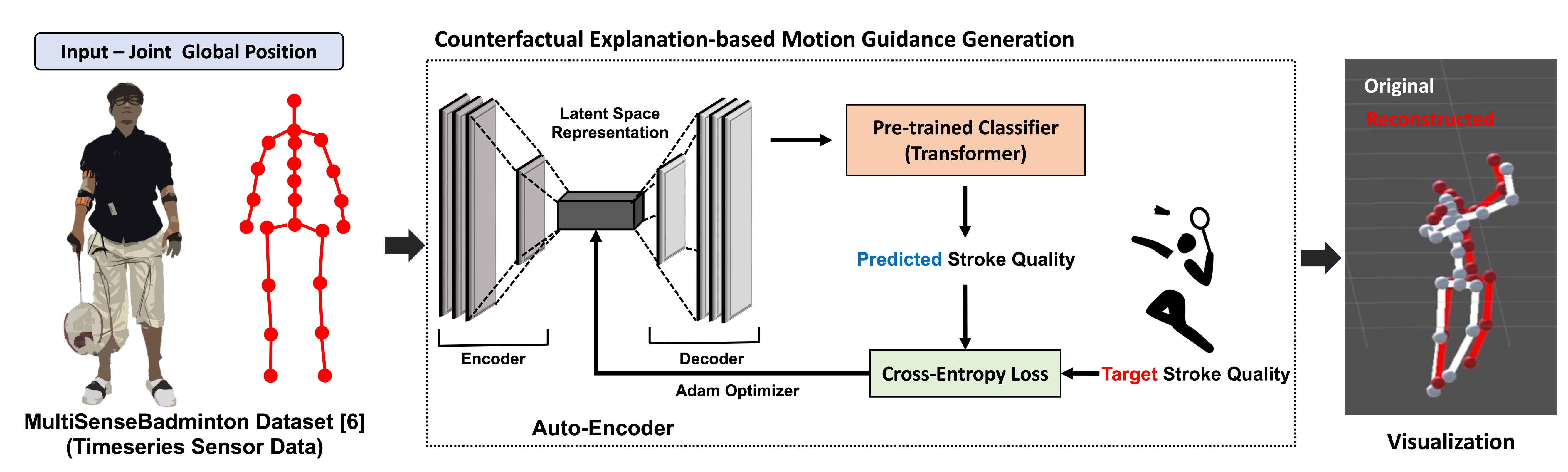}
\caption{Motion guidance generation framework}
\label{fig:framework}
\end{figure*}

\begin{abstract}
This study proposes a framework for enhancing the stroke quality of badminton players by generating personalized motion guides, utilizing a multimodal wearable dataset. These guides are based on counterfactual algorithms and aim to reduce the performance gap between novice and expert players. Our approach provides joint-level guidance through visualizable data to assist players in improving their movements without requiring expert knowledge. The method was evaluated against a traditional algorithm using metrics to assess validity, proximity, and plausibility, including arithmetic measures and motion-specific evaluation metrics. Our evaluation demonstrates that the proposed framework can generate motions that maintain the essence of original movements while enhancing stroke quality, providing closer guidance than direct expert motion replication. The results highlight the potential of our approach for creating personalized sports motion guides by generating counterfactual motion guidance for arbitrary input motion samples of badminton strokes.
\end{abstract}

\section{INTRODUCTION}

Learning sports skills often involves observing and imitating expert athletes’ movements \cite{liao2022ai}. However, novices face challenges in determining the quality of their actions and in improving their skills without the guidance or prior knowledge of a coach. The scarcity of guidance for reproducing the exact movements of an expert or a successful sports action can lead to difficulties in fine-tuning movements \cite{liao2022ai}. 

In particular, racket sports pose additional training challenges owing to the rapid nature of the actions, which requires players to correct their movements based on feedback received. Given the fast-paced nature of stroke actions in racket sports, it is often difficult for experts to holistically capture this feedback.

In this regard, the use of multiple wearable sensor data has made it possible to analyze how athletes coordinate their movements without prior knowledge \cite{qiu2022multi}. These sensors allow for the capture of motion from multiple viewpoints and at a microsecond-level time resolution. Hence, these sensors provide quantitative evidence of motion quality and areas for improvement. 

In addition, advances in AI technology have revolutionized sports training. AI can not only predict ball trajectories and analyze game strategies \cite{seong2023team}, but also assess individual movements and provide personalized training programs \cite{ghosh2022decoach}. In this context, we propose a framework for generating visualizable joint-level motion guidance by leveraging counterfactual (CF) explanations derived from latent space-level differences in the motions of novices and experts \textbf{(see Figure \ref{fig:framework})}. Counterfactual explanations refer to the understanding of how a given outcome would change if certain aspects of the initial conditions were altered, providing a "what if" analysis that can illuminate the path from novice to expert performance \cite{wachter2017counterfactual}. By integrating counterfactual explanations with multimodal wearable sensor data, the framework provides feedback on how novice players modify their movements to achieve expert-level performance.

\section{METHODS}

\subsection{Wearable Sensor-based Badminton Dataset}

We used the MultiSenseBadminton dataset \cite{seong2024multisensebadminton}, which contains 7,763 badminton swing motions and annotations regarding the performance of each swing. The dataset comprised 25 badminton players with varying skill levels, including 12 beginners, 8 intermediate players, and 5 experts. For each player, data were collected for two swing types: forehand high clear drive and backhand drive. The dataset was collected while participants were equipped with multiple wearable sensors, allowing for the simultaneous measurement of eye tracking, muscle activity, foot pressure, and joint data. The dataset also provides annotations for each swing, including the ball landing location, hit point, and stroke quality. We used the joint global position data as our input and the stroke quality as our target annotation to relate swing mechanics and stroke quality for motion guidance generation.

\begin{algorithm}
\caption{LatentCF-based motion guidance generation}
\small
\label{alg:cf_algorithm}
\begin{algorithmic}[1]
\State \textbf{input:} A time series motion sample $x$, target stroke quality class $y'$, learning rate $\alpha$, maximum iteration $\textit{max\_iter}$, pre-trained classifier $C$, pre-trained autoencoder $AE$
\State \textbf{Output:} A generated counterfactual motion $x'$ with desired target class $y'$

\State $z \gets \text{AE-Encode}(x)$
\State $y_{pred} \gets \text{C}(\text{AE-Decode}(z))$
\State $\textit{loss} \gets \text{CrossEntropyLoss}(y_{pred} - y')$
\State $\textit{iter} \gets 0$

\While{$y_{pred} < \tau \land \textit{iter} < \textit{max\_iter}$}
    \State $z \gets \text{AdamOptimize}(z, \textit{loss}, \alpha)$
    \State $y_{pred} \gets \text{C}(\text{AE-Decode}(z))$
    \State $\textit{loss} \gets \text{CrossEntropyLoss}(y_{pred} - y')$
    \State $\textit{iter} \gets \textit{iter} + 1$
\EndWhile
\State $x' \gets \text{AE-Decode}(z)$
\State \textbf{return $x'$}

\end{algorithmic}
\end{algorithm}

\subsection{Motion Guidance Generation Framework}

This framework aims to generate improved full-body motion guidance trajectories, guiding users from their current motion towards expert-like motions without direct imitation, by leveraging a network trained on expert performances. Our framework for generating motion guides consists of three components: 1) a \textbf{pretrained classifier} that utilizes badminton players’ joint global position data to predict stroke quality, 2) a \textbf{model explainer} that applies a CF algorithm to generate motion guidance based on the original motion data, and 3) a \textbf{motion visualizer}. The model explainer transforms the input data into a latent space using an autoencoder and adjusts it based on discrepancies between the model's predicted stroke quality and the target stroke quality, and a decoder is used to restore the adjusted data (see \textbf{Algorithm \ref{alg:cf_algorithm}}). Finally, the motion visualizer implemented in Unity was used to compare the original and CF-generated data.

\section{EXPERIMENT AND ANALYSIS}

\subsection{Model Training Results} 
We built a model to predict stroke quality based on wearable sensor data, and trained it on people across different skill levels. We selected the Conv1D, ConvLSTM, LSTM, and Transformer models. A baseline model was also established to predict the majority class for comparison.

In terms of metrics for evaluating the model performance, we utilized accuracy, balanced accuracy, and the F1 score. Each model underwent a fivefold cross-validation process. Additionally, data augmentation was performed on the training dataset to enhance the model training and generalization.

After reviewing the outcomes of our model training, as shown in \textbf{Table \ref{tab:Classification Results}}, we chose the transformer model for our final implementation. It consistently achieved higher scores across all evaluated metrics than did the other models.

\begin{table}[htb!]
\caption{Evaluation Result of Stroke Quality Classifier}
\label{tab:Classification Results}
\centering
{\small
\begin{tabular}{@{}cccc@{}}
\toprule
 \multirow{2}{*}{\textbf{Model}} & \multicolumn{3}{c}{\textbf{Fivefold Forehand Clear Results}} \\
\cmidrule(r){2-4}
  & \textbf{Acc$_{avg}$ (SD)} & \textbf{BalAcc$_{avg}$ (SD)} & \textbf{F1$_{avg}$ (SD)} \\
\midrule

 Transformer & \textbf{85.12 (1.62)} & \textbf{85.67 (1.57)} & \textbf{85.22 (1.60)} \\
   Conv1D & 79.97 (2.04) & 76.82 (3.43) & 78.94 (2.70)  \\
  ConvLSTM & 80.83 (4.15) & 78.41 (6.65) & 79.74 (5.51)  \\
  LSTM & 84.29 (1.78) & 84.84 (1.72) & 84.40 (1.77) \\
  Baseline & 59.17 (2.25) & 50.00 (0.00) & 44.01 (2.72) \\
 
\midrule
 \multirow{2}{*}{\textbf{Model}} & \multicolumn{3}{c}{\textbf{Fivefold Backhand Drive Results}} \\
\cmidrule(r){2-4}
  & \textbf{Acc$_{avg}$ (SD)} & \textbf{BalAcc$_{avg}$ (SD)} & \textbf{F1$_{avg}$ (SD)} \\

\midrule
 
Transformer & 87.89 (0.89) & \textbf{88.93 (0.76)} & \textbf{88.22 (0.87)} \\  
Conv1D & 87.53 (3.03) & 88.16 (1.50) & 87.84 (2.85)  \\
  ConvLSTM & 85.80 (7.57) & 85.92 (3.57) & 86.03 (7.16) \\  
  LSTM & \textbf{87.92 (0.77)} & 88.15 (1.13) & 88.19 (0.74) \\
  Baseline & 70.77 (2.03) & 50.00 (0.00) & 58.68 (2.68) \\

\bottomrule
\end{tabular}
}

\end{table}

\subsection{Motion Guidance Generation Results} 

We compared motion guides created by our LatentCF algorithm against those generated by a 1-nearest neighbor (1NN) baseline, calculated using L1 distance, L2 distance, and Dynamic Time Warping (DTW) \cite{delaney2021instance, wang2021learning, wang2021counterfactual}, utilizing standard CF-specific and motion-specific metrics such as validity (accuracy of the CFs in predicting targets) \cite{wang2021learning}, proximity (measured by L1, L2, and L$_{\text{inf}}$ distances) \cite{delaney2021instance, wang2021learning}, plausibility assessed via local outlier factor (LOF), isolation forest (IF), and one class support vector machine (OCSVM) methods \cite{delaney2021instance, wang2021counterfactual}, closeness determined via DTW \cite{wang2021learning}, Fréchet motion distances (FMD) and Fréchet pose distances (FPD) \cite{au2022choreograph}. For the 1NN-based CF example, we select the nearest example predicted to have the target stroke quality among the neighbors obtained through 1NN, utilizing raw data for this purpose. To evaluate the performance of each CF generation algorithm, we randomly selected 100 instances of each stroke type, calculated their metrics, and averaged them. Generated CF algorithms were visualized as ghost avatars overlaid on the user's avatar, with examples at https://youtu.be/o7bDM5yRbtw.

\begin{table}[htb!]
\caption{Evaluation results of Counterfactual Motion Guidance}
\label{tab:CF Results}
\centering
{\scriptsize
\begin{tabular}{ccccc}
\toprule
 \multirow{2}{*}{\textbf{Metric}} & \multicolumn{4}{c}{\textbf{Forehand Clear}}  \\
\cmidrule(r){2-5}
  & \textbf{Ours (SD)}  & \textbf{1NN$_{L1}$ (SD)} & \textbf{1NN$_{L2}$ (SD)}  & \textbf{1NN$_{DTW}$ (SD)} \\
\midrule

 Validity ${\uparrow}$ & \textbf{1}  & \textbf{1}  & \textbf{1}  & \textbf{1} \\
 
  L1 ${\downarrow}$& \textbf{3.41 (1.07)} & 5.43 (1.63) & 5.51 (1.62) &  5.54 (1.88)  \\

   L2 ${\downarrow}$& \textbf{0.58 (0.18)}& 0.91 (0.26) & 0.91 (0.26)   &  0.92 (0.28) \\

   L$_{inf}$ ${\downarrow}$& \textbf{0.23 (0.06)} & 0.35 (0.09) & 0.34 (0.09)   & 0.35 (0.09)\\
  
  LOF ${\downarrow}$& \textbf{0.36 (0.26)} & 0.63 (0.30) & 0.62 (0.30)  &  0.62 (0.29) \\
  
  IF ${\downarrow}$& \textbf{0.91 (0.13)} & 0.97 (0.07) & 0.97 (0.06)  & 0.97 (0.08) \\

  OCSVM ${\downarrow}$& \textbf{0.55 (0.21)} & 0.68 (0.22) & 0.69 (0.21)  & 0.69 (0.22) \\

  DTW ${\downarrow}$ & \textbf{3.39 (1.16)} & 5.36 (1.66) & 5.42 (1.65)   &  5.41 (1.75) \\

  FPD ${\downarrow}$ & \textbf{0.33 (0.37)}   & 0.71 (0.56) & 0.70 (0.56)  & 0.69 (0.44) \\

  FMD ${\downarrow}$ & \textbf{1.02 (1.13)} & 2.16 (1.71) & 2.14 (1.71)  & 2.09 (1.34) \\

\midrule

   \multirow{2}{*}{\textbf{Metric}} &   \multicolumn{4}{c}{\textbf{Backhand Drive}} \\
\cmidrule(r){2-5}
  & \textbf{Ours (SD)}  & \textbf{1NN$_{L1}$ (SD)} & \textbf{1NN$_{L2}$ (SD)}  & \textbf{1NN$_{DTW}$ (SD)} \\
\midrule

 Validity ${\uparrow}$ & \textbf{1} & \textbf{1} & \textbf{1} & \textbf{1}  \\
 
  L1 ${\downarrow}$& \textbf{2.99 (0.98)}  & 3.92 (1.14) & 3.97 (1.15) & 3.97 (1.16) \\

   L2 ${\downarrow}$& \textbf{0.51 (0.16)} & 0.67 (0.19) & 0.67 (0.19) & 0.67 (0.19) \\

   L$_{inf}$ ${\downarrow}$&  \textbf{0.21 (0.06)}& 0.28 (0.08) & 0.26 (0.08) &  0.27 (0.08) \\
  
  LOF ${\downarrow}$& \textbf{0.44 (0.21)} & 0.63 (0.26) & 0.59 (0.27) & 0.61 (0.26)  \\
  
  IF ${\downarrow}$&  0.99 (0.03) & 0.99 (0.05) & \textbf{0.98 (0.07)} & \textbf{0.98 (0.07)} \\

  OCSVM ${\downarrow}$& \textbf{0.61 (0.16)} & 0.78 (0.21) &  0.77 (0.22) & 0.77 (0.21) \\

  DTW ${\downarrow}$ & \textbf{3.02 (1.06)}& 3.87 (1.10) & 3.92 (1.11) & 3.91 (1.11)  \\

  FPD ${\downarrow}$ &  \textbf{0.28 (0.21)} & 0.44 (0.25) & 0.42 (0.25) & 0.44 (0.25) \\

  FMD ${\downarrow}$ & \textbf{0.85 (0.65)} & 1.33 (0.75) & 1.28 (0.75) &  1.32 (0.76) \\

\bottomrule
\end{tabular}
}

\end{table}

The summarized results in \textbf{Table \ref{tab:CF Results}}, show our CF method outperforming the 1NN baselines, significantly reducing the L1, L2, and L$_{\text{inf}}$ distances and closely mirroring the original movements. Our approach excels in temporal and sequential alignment, as shown by the DTW, and it demonstrates high validity, thereby affirming its accuracy for motion generation. With plausibility confirmed via LOF, ISO, and OCSVM as well as the motion-specific metrics provided by FPD and FMD, our method enhances stroke quality. 


\section{FUTURE WORKS AND CONCLUSIONS}

This study presents a method using CF algorithms to generate badminton guide motions from player movements. In future work, we will use multitask learning for dual assessment of stroke quality and direction, and include user evaluations to validate real-world applicability. Specifically, we plan to incorporate various visual and haptic systems for motion guidance, such as Electrical Muscle Stimulation \cite{hwang2024ergopulse}, electrical and vibration feedback \cite{hwang2023enhancing, hwang2023electrical, hwang2022reves}, and diverse visual cues \cite{wu2020vr, wu2019vizski}, to enhance the effectiveness of our framework. Additionally, we plan to explore the applications of our multimodal data-based guide motion generation framework in robotics, leveraging its potential for skill transfer and imitation learning.

\addtolength{\textheight}{-12cm}

\section*{ACKNOWLEDGMENT}

This work was supported by the GIST-MIT Research Collaboration grant funded by the GIST in 2024. We appreciate the high-performance GPU computing support of HPC-AI Open Infrastructure via GIST SCENT.

\bibliographystyle{IEEEtran}
\bibliography{ref}

\begin{thebibliography}{10}
\providecommand{\url}[1]{#1}
\csname url@samestyle\endcsname
\providecommand{\newblock}{\relax}
\providecommand{\bibinfo}[2]{#2}
\providecommand{\BIBentrySTDinterwordspacing}{\spaceskip=0pt\relax}
\providecommand{\BIBentryALTinterwordstretchfactor}{4}
\providecommand{\BIBentryALTinterwordspacing}{\spaceskip=\fontdimen2\font plus
\BIBentryALTinterwordstretchfactor\fontdimen3\font minus \fontdimen4\font\relax}
\providecommand{\BIBforeignlanguage}[2]{{%
\expandafter\ifx\csname l@#1\endcsname\relax
\typeout{** WARNING: IEEEtran.bst: No hyphenation pattern has been}%
\typeout{** loaded for the language `#1'. Using the pattern for}%
\typeout{** the default language instead.}%
\else
\language=\csname l@#1\endcsname
\fi
#2}}
\providecommand{\BIBdecl}{\relax}
\BIBdecl

\bibitem{liao2022ai}
C.-C. Liao, D.-H. Hwang, and H.~Koike, ``Ai golf: Golf swing analysis tool for self-training,'' \emph{IEEE Access}, vol.~10, pp. 106\,286--106\,295, 2022.

\bibitem{qiu2022multi}
S.~Qiu, H.~Zhao, N.~Jiang, Z.~Wang, L.~Liu, Y.~An, H.~Zhao, X.~Miao, R.~Liu, and G.~Fortino, ``Multi-sensor information fusion based on machine learning for real applications in human activity recognition: State-of-the-art and research challenges,'' \emph{Information Fusion}, vol.~80, pp. 241--265, 2022.

\bibitem{seong2023team}
M.~Seong, J.~Oh, and S.~Kim, ``Team badminseok at ijcai coachai badminton challenge 2023: Multi-layer multi-input transformer network (mulminet) with weighted loss,'' \emph{arXiv preprint arXiv:2307.08262}, 2023.

\bibitem{ghosh2022decoach}
I.~Ghosh, S.~R. Ramamurthy, A.~Chakma, and N.~Roy, ``Decoach: Deep learning-based coaching for badminton player assessment,'' \emph{Pervasive and Mobile Computing}, vol.~83, p. 101608, 2022.

\bibitem{wachter2017counterfactual}
S.~Wachter, B.~Mittelstadt, and C.~Russell, ``Counterfactual explanations without opening the black box: Automated decisions and the gdpr,'' \emph{Harv. JL \& Tech.}, vol.~31, p. 841, 2017.

\bibitem{seong2024multisensebadminton}
M.~Seong, G.~Kim, D.~Yeo, Y.~Kang, H.~Yang, J.~DelPreto, W.~Matusik, D.~Rus, and S.~Kim, ``Multisensebadminton: Wearable sensor--based biomechanical dataset for evaluation of badminton performance,'' \emph{Scientific Data}, vol.~11, no.~1, p. 343, 2024.

\bibitem{delaney2021instance}
E.~Delaney, D.~Greene, and M.~T. Keane, ``Instance-based counterfactual explanations for time series classification,'' in \emph{International conference on case-based reasoning}.\hskip 1em plus 0.5em minus 0.4em\relax Springer, 2021, pp. 32--47.

\bibitem{wang2021learning}
Z.~Wang, I.~Samsten, R.~Mochaourab, and P.~Papapetrou, ``Learning time series counterfactuals via latent space representations,'' in \emph{Discovery Science: 24th International Conference, DS 2021, Halifax, NS, Canada, October 11--13, 2021, Proceedings 24}.\hskip 1em plus 0.5em minus 0.4em\relax Springer, 2021, pp. 369--384.

\bibitem{wang2021counterfactual}
Z.~Wang, I.~Samsten, and P.~Papapetrou, ``Counterfactual explanations for survival prediction of cardiovascular icu patients,'' in \emph{Artificial Intelligence in Medicine: 19th International Conference on Artificial Intelligence in Medicine, AIME 2021, Virtual Event, June 15--18, 2021, Proceedings}.\hskip 1em plus 0.5em minus 0.4em\relax Springer, 2021, pp. 338--348.

\bibitem{au2022choreograph}
H.~Y. Au, J.~Chen, J.~Jiang, and Y.~Guo, ``Choreograph: Music-conditioned automatic dance choreography over a style and tempo consistent dynamic graph,'' in \emph{Proceedings of the 30th ACM International Conference on Multimedia}, 2022, pp. 3917--3925.

\bibitem{hwang2024ergopulse}
S.~Hwang, J.~Oh, S.~Kang, M.~Seong, A.~I. A.~M. Elsharkawy, and S.~Kim, ``Ergopulse: Electrifying your lower body with biomechanical simulation-based electrical muscle stimulation haptic system in virtual reality,'' in \emph{Proceedings of the CHI Conference on Human Factors in Computing Systems}, 2024, pp. 1--21.

\bibitem{hwang2023enhancing}
S.~Hwang, Y.~Kim, Y.~Seo, and S.~Kim, ``Enhancing seamless walking in virtual reality: Application of bone-conduction vibration in redirected walking,'' in \emph{2023 IEEE International Symposium on Mixed and Augmented Reality (ISMAR)}.\hskip 1em plus 0.5em minus 0.4em\relax IEEE, 2023, pp. 1181--1190.

\bibitem{hwang2023electrical}
S.~Hwang, J.~Lee, Y.~Kim, Y.~Seo, and S.~Kim, ``Electrical, vibrational, and cooling stimuli-based redirected walking: Comparison of various vestibular stimulation-based redirected walking systems,'' in \emph{Proceedings of the 2023 CHI Conference on Human Factors in Computing Systems}, 2023, pp. 1--18.

\bibitem{hwang2022reves}
S.~Hwang, J.~Lee, Y.~Kim, and S.~Kim, ``Reves: Redirection enhancement using four-pole vestibular electrode stimulation,'' in \emph{CHI Conference on Human Factors in Computing Systems Extended Abstracts}, 2022, pp. 1--7.

\bibitem{wu2020vr}
E.~Wu, T.~Nozawa, F.~Perteneder, and H.~Koike, ``Vr alpine ski training augmentation using visual cues of leading skier,'' in \emph{Proceedings of the IEEE/CVF Conference on Computer Vision and Pattern Recognition Workshops}, 2020, pp. 878--879.

\bibitem{wu2019vizski}
E.~Wu, F.~Perteneder, H.~Koike, and T.~Nozawa, ``How to vizski: Visualizing captured skier motion in a vr ski training simulator,'' in \emph{Proceedings of the 17th International Conference on Virtual-Reality Continuum and Its Applications in Industry}, 2019, pp. 1--9.

\end{thebibliography}

\end{document}